\documentclass[10pt,twocolumn,showkeys,nofootinbib]{revtex4-1}

\long\def\dontignore#1{#1}
\makeatletter

\long\def\ignoreflag{\@makeother\{\@makeother\}\xignore}
\long\def\xignore#1\dontignore#2{\catcode`\{\@ne
\catcode`\}\tw@
\afterassignment\xxdontignore\toks@\bgroup}
\long\def\xxdontignore{\the\toks@\ignoreflag}
\makeatother
\let\ignoreflag\relax

\usepackage{nicefrac}
\usepackage{tikz}
\usetikzlibrary{calc}
\usetikzlibrary{positioning}
\usetikzlibrary{decorations,positioning,shapes}
\usetikzlibrary{decorations.pathmorphing}

\usetikzlibrary{decorations.shapes}
\usetikzlibrary{decorations.markings}

\tikzstyle{vecArrow} = [thick, decoration={markings,mark=at position
   1 with {\arrow[semithick]{open triangle 60}}},
   double distance=1.4pt, shorten >= 5.5pt,
   preaction = {decorate},
   postaction = {draw,line width=1.4pt, white,shorten >= 4.5pt}]
\tikzstyle{innerWhite} = [semithick, white,line width=1.4pt, shorten >= 4.5pt]

\usepackage{natbib}
\usepackage{paralist}

\usepackage{amsmath}
\usepackage{amssymb}
\usepackage{import}

\usepackage{subfigure}

\usepackage{textcomp}
\usepackage[version=3]{mhchem}
\usepackage[utf8]{inputenc}

\usepackage{eurosym}

\usepackage{rotating}
\usepackage{multirow}

\definecolor{contour}{RGB}{16,16,127}
\definecolor{branchcut}{RGB}{255,63,0}

\newcommand{\manualcite}[1]{}

\newcommand{\bra}[1]{\left\langle #1 \right\vert}

\newcommand{\matrixel}[3]{\left< #1 \vphantom{#2#3}\right| #2 \left| #3 \vphantom{#1#2}\right>}
\newcommand{\braket}[2]{\left< #1 \vphantom{#2}\right| \left. #2 \vphantom{#1}\right>}

\newcommand{\unit}[1]{
\,\mathrm{#1}
}

\newcommand{\di}{\mathrm{d}}

\newcommand{\e}{\,\mathrm{e}}
\newcommand{\erf}{\,\mathrm{erf}}

\newcommand{\op}[1]{\mathcal{#1}}

\newcommand{\order}[1]{\mathcal{O}(#1)}

\renewcommand{\imath}{{\mathrm i}}
\newcommand{\numberset}[1]{{\mathbb #1}}

\makeatletter\newcommand{\restoretitle}[1]{
\newcommand{\GNUPLOTspecial}{  \@sanitize\catcode`\%=14\relax\special}  {\GNUPLOTspecial{"
SDict begin [
  /Title (#1)
  /Subject (#1 Application)
  /Creator (LaTeX)
  /Author (Werner Koch)
  /CreationDate (Mon Mar  5 10:17:40 2012)
  /DOCINFO pdfmark
end
}}}
\makeatother

\renewcommand{\theequation}{\arabic{equation}}

\usepackage[T1]{fontenc}

\DeclareUnicodeCharacter{2010}{-}

\newcommand{\mysection}[1]{\paragraph*{#1}}
\newcommand{\mysubsection}[1]{\paragraph*{#1}}

\providecommand{\keywords}[1]{\textbf{\textit{Index terms---}} #1}
\begin{document}
\ignoreflag

\date{\today}
\title{A Three-step Model of High Harmonic Generation using Complex Classical Trajectories}
\author{Werner Koch}
\affiliation{Weizmann Institute of Science, Rehovot, Israel}
\author{David J. Tannor}
\affiliation{Weizmann Institute of Science, Rehovot, Israel}

\begin{abstract}
We present a new trajectory formulation of high harmonic generation that treats classically allowed and classically forbidden processes within a single dynamical framework. Complex trajectories orbit the nucleus, producing the stationary Coulomb ground state. When the field is turned on, these complex trajectories continue their motion in the field-dressed Coulomb potential and therefore tunnel ionization, unbound evolution and recollision are described within a single, seamless framework. The new formulation can bring mechanistic understanding to a broad range of strong field physics effects.

\end{abstract}
\maketitle

\mysection{Introduction}

The strong field dynamics of Coulombic systems has been at the center of attention of the atomic, molecular and optical physics communities for decades. Particularly fascinating are the effects of above threshold ionization (ATI) \cite{agostini_free-free_1979} and high harmonic generation (HHG) \cite{mcpherson_studies_1987,ferray_multiple-harmonic_1988,li_multiple-harmonic_1989}. Besides being of interest for its fundamental physics, HHG is a powerful tool for generating XUV pulses that can be used to further investigate atomic and molecular quantum dynamics.

Above threshold ionization and high harmonic generation are intrinsically quantum mechanical phenomena, and where numerical evaluation of the time dependent Schrödinger Equation (TDSE) is possible it faithfully reproduces the experimentally observed quantities. However, numerical simulation does not always provide mechanistic understanding, and hence there is great interest in trajectory-based classical and semi-classical models to describe these phenomena \cite{protopapas_recollisions_1996,van_de_sand_irregular_1999}.

Consider the three-step model of HHG introduced by Corkum \cite{corkum_plasma_1993} (see also \cite{kulander_dynamics_1993}). In the three-step model, the HHG process is decomposed as follows: 1) tunnel ionization 2) free propagation in the presence of the field, and 3) recollision and recombination producing HHG emission \cite{corkum_plasma_1993,lewenstein_theory_1994}.
It is found that classical models are accurate for the second step of the three-step model -- the free propagation in the presence of the field.
But real-valued trajectories cannot describe the tunnel ionization that takes place in the first step.
Moreover, real-valued trajectories in the Coulomb system cannot describe a stationary initial wavefunction because all trajectories fall into the nucleus. This undermines a consistent formulation for sampling initial conditions from the stationary wavefunction \footnote{This is true for 1D models and for multi electron systems. Two and three dimensional single electron models do not have this problem.}.
As a result of these limitations, classical-based models have generally chosen initial conditions on the outside of the field-induced Coulomb barrier, with the choice of initial momentum zero.

In this paper we show that complex-valued classical trajectories can overcome all of the above problems, accounting for tunnel ionization, laser dressed electron propagation and recollision, all within a single consistent framework.
For the purposes of presentation we define a modified three-step model:
1) Construction of the stationary ground state Coulomb wavefunction using complex trajectories. It is difficult to overstate how non-trivial this step is, as will become clear below;
2) Evolution of the complex trajectories in the presence of the Coulomb potential plus time-dependent strong field.
Note that in this step tunneling ionization, free evolution and recombination are all treated in a seamless, unified way.
3) Construction of the high harmonic spectrum from the Fourier transform of the time-dependent dipole acceleration. At every stage, the time-dependent wavefunction is reconstructed and shown to be in semi-quantitative agreement with the exact quantum result.

This is by no means the first attempt to use complex trajectories to describe strong field ionization \cite{perelomov_ionization_1966,perelomov_ionization_1967,perelomov_ionization_1967-1,keldysh_ionization_1964,faisal_collision_1973,reiss_effect_1980,salieres_feynmans_2001,shafir_resolving_2012,kaushal_nonadiabatic_2013,torlina_interpreting_2015,pisanty_slalom_2016,kaushal_looking_2018-1} but we believe this is the first treatment where all portions of the three-step model are described within a single, consistent trajectory framework, sampled from an initial Coulomb eigenstate.
In some sense it is the completion of the program of the classic papers by  Perelomov, Popov, and Terent'ev (PPT) \cite{perelomov_ionization_1966}, Corkum \cite{corkum_plasma_1993} and Lewenstein et al. \cite{lewenstein_theory_1994}.

\mysection{Theory}
The initial state of the system $\Psi_0(x)$, is analytically continued into the complex plane. This defines a complex-valued action $S_0(x)$
\begin{align}
  \label{eq:action}
\Psi_0(x)=\e^{\frac{\imath}{\hbar}S_0(x)} \quad S_0(x),x\in\numberset{C}\,.
\end{align}
A manifold of complex-valued initial conditions $q_0,p_0\in \numberset{C}$ is defined through the relation
\begin{align}
\label{eq:manifold}
  p_0(q_0)=\left.\frac{\partial S_0(x)}{\partial x} \right\vert_{x=q_0}\,,
\end{align}
with a stability parameter $\alpha_0(q_0)=\left.\frac{\partial^2 S_0(x)}{\partial x^2} \right\vert_{x=q_0}$.
The complex trajectories are then propagated by integrating the equations of motion (EOM)
\begin{align}
\label{eq:EOM}
  \begin{split}
    \dot{q} &= \frac{1}{m} p \\
    \dot{p} &= -V'(q)\\
    \dot{S} &= \frac{1}{2m}p^2 -V(q) + \frac{\imath\hbar}{2m}\alpha\\
    \dot{\alpha}&=-V''(q)-\frac{1}{m}\alpha^2
  \end{split}
\end{align}
where $m$ is the mass of the system.
The complex action $S=S^{\rm cl}+S^{\rm qm}$, where $S^{\rm cl}=\int \di t \frac{p^2}{2m}-V(q)$ and $S^{\rm qm}=\int\di t \frac{\imath\hbar}{2m}\alpha$.

In the final value coherent state propagator (FINCO) method \cite{zamstein_communication:_2014}, each trajectory is associated with a matrix element of the propagator in the coherent state representation $\matrixel{g_{\gamma}(\xi(q_0))}{\op{U}}{\Psi_0}$. Equations ~\eqref{eq:EOM} subject to initial conditions \eqref{eq:manifold} determine the complex coordinate $q_t(q_0)$ and momentum $p_t(q_0)$ at time $t$. This in turn provides a coherent state label $\xi(q_0)$ for the bra Gaussian $\bra{g_{\gamma}(\xi(q_0))}$ through the relation
\begin{align}
  \label{eq:HHtrafo}
2\gamma q_t(q_0)-\imath p_t(q_0)&=2\gamma \bar{q}_t(q_0)-\imath \bar{p}_t(q_0) \equiv \xi(q_0)
\end{align}
where $q_t, p_t \in\numberset{C}$, $\bar{q}_t, \bar{p}_t,\gamma \in\numberset{R}$ \cite{huber_generalized_1987}.
The coherent state matrix element is given by $\matrixel{g_{\gamma}(\xi(q_0))}{\op{U}}{\Psi_0}=\phi(q_0)\e^{\sigma(q_0)}$, where the exponent $\sigma(q_0)$ has the form
\begin{align}
\label{eq:sigma}
\sigma(q_0)&=\frac{\imath}{\hbar} S_{t}^{\rm cl}(q_0)+\frac{1}{4\gamma\hbar^2}p_t^2(q_0)-\frac{1}{4\gamma}\bigl(\Im \xi(q_0)\bigr)^2\,.
\end{align}
Associated with this bra Gaussian is a ket Gaussian centered at $\xi^{\ast}=2\gamma\bar{q}_t(q_0)+\imath\bar{p}_t(q_0)$,
\begin{align}
  \label{eq:gaussian}
\braket{x}{g_{\gamma}(\xi(q_0))}
= \left(\tfrac{2\gamma}{\pi}\right)^{\frac{1}{4}}\e^{ -\gamma\left[ x-\bar{q}_t(q_0)\right]^{2}
-\imath\bar{p}_t(q_0)\left[ x-\bar{q}_t(q_0)\right]}\,.
\end{align}
The time evolved wavefunction $\Psi_t(x)$ is represented in a basis of these ket Gaussians
\begin{align}
\label{eq:finco}
\hspace{-0.25em}\Psi_t(x)=\int\frac{-\di q_0}{4\pi\gamma}|J(q_0)| \braket{x}{g_{\gamma}(\xi(q_0))}\matrixel{g_{\gamma}(\xi(q_0))}{\op{U}}{\Psi_0}\,.
\end{align}

The Jacobian
$|J(q_0)|=\left|\frac{\di \xi(q_0)}{\di q_0}\right|^2$
and prefactor
$\phi(q_0)=\left(8\gamma\pi\right)^{\frac{1}{4}}\left[\frac{\di \xi(q_0)}{\di q_0}\right]^{-\frac{1}{2}}$,
where
\begin{align}
  \label{eq:DxiDnu}
\frac{\di \xi(q_0)}{\di q_0}
=\left(2\gamma M_{qp}{-}\imath M_{pp}\right)\alpha_0(q_0){+}\left(2\gamma M_{qq}{-}\imath M_{pq}\right)\,.
\end{align}
The $M_{ab}$ are stability matrix elements defined by $M_{ab}=\frac{\partial a_t(q_0)}{\partial b_0(q_0)}$ with $a,b\in\{p,q\}$. EOM for the $M_{ab}$ are given in App.~\ref{sec:stability}.

\mysubsection{Complex time}
When the EOM of the previous section are applied to the Coulomb system, the methodology fails at the first step: no meaningful reconstruction of the ground state eigenfunction is obtained. Real-valued bound state initial conditions all fall into the Coulomb singularity.
Strictly complex-valued trajectories, on the other hand, diverge, as the complex Coulomb potential is repulsive everywhere except on the real axis. Hence a stable ground state can not be constructed from either real or complex trajectories.

The resolution to this conundrum comes from allowing time to be complex. The complex time plane has singularities -- points where the classical momentum goes to infinity in a finite time. Paradoxically, these singularities, rather than being a nuisance, provide the solution to the problem by changing the topology of the trajectories. For instance, it is known that circumnavigating a time-singularity in barrier scattering problems can transform a transmitted trajectory into a reflected trajectory \cite{doll_complexvalued_1973,kay_time-dependent_2013,petersen_complex_2014}. In the Coulomb system, circumnavigating complex time singularities leads to complex trajectories that orbit the nucleus. Technically, these singularities cause the solution $q_t(q_0),p_t(q_0)$ to be multivalued, but the motion is continuous when viewed as evolving on a multi-sheeted Riemann surface. Motion on these sheets may correspond to either classically allowed or classically forbidden processes, depending on the trajectory initial conditions.

\begin{figure}[htp]
  \centering
\includegraphics[width=\columnwidth]{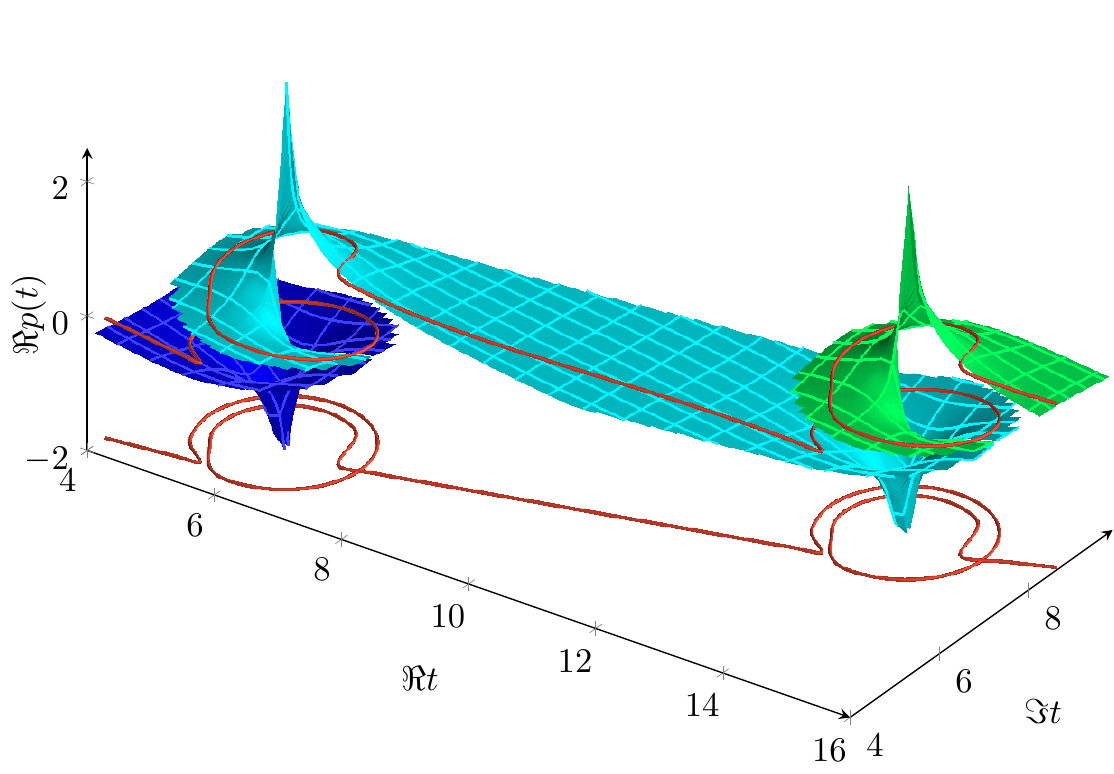}
\caption{\label{fig:riemann}Riemann surface of the real part of the complex momentum as a function of complex time $p(t)$.
Each sheet is indicated by a different color.
Two singularities are visible.
The integration contour, shown in red, loops around these singularities, thus passing to the next sheet. For clarity, the integration contour is shown on the base of the plot as well as superimposed on the Riemann surfaces.}
\end{figure}

To understand more fully the influence of circumnavigating a singularity in time, we note that the singularities in complex time arise from singularities in complex position \cite{doll_complexvalued_1973, kay_time-dependent_2013}. In Ref.~\cite{koch_multivalued_2018-1} a full calculus was introduced relating the asymptotic approach towards a singularity of the potential to the singularities in complex time. As shown in \cite{koch_multivalued_2018-1}, the Coulomb singularity at the origin gives rise to three Riemann sheets for each singularity encounter in time.  Controlled circumnavigation of the singularity in time leads to complex trajectories that cycle through these three Riemann sheets, corresponding to trajectories that orbit the nucleus (Fig. \ref{fig:riemann}).

Associated with the singularities in complex position there are always caustics, although the mathematical connection between these two is yet to be discovered. These caustics lead to divergence of a portion of the complex trajectories, those that lie beyond what are known as Stokes lines.  A rigorous procedure for removing these unphysical trajectories was introduced in Ref.~\cite{koch_communication:_2018}.
\mysection{The three-step model. Step 1: Constructing the ground state of the Coulomb system}

The first step is to describe the stationary initial state of the Coulomb system in terms of orbiting trajectories. Consider the anti-symmetric, Cartesian 1D Coulomb ground state $\Psi_0(x)=2x\e^{-\vert x\vert}$.
The strategy is to let these trajectories orbit the nucleus freely until the field is turned on and then seamlessly continue their propagation in the presence of the field.

A complex time contour that loops around a singularity time passes onto a second Riemann sheet where the corresponding trajectory orbits the nucleus. On this alternate sheet, an additional singularity time is present which is in turn circumnavigated by a loop contour.
Continuing in this way, each loop corresponds to a trajectory passing from one sheet of the Riemann surface to another, followed by propagation on that new sheet until the next singularity encounter.
Figure~\ref{fig:riemann} shows this transition between sheets for two successive singularity times.

\begin{figure}[htp]
  \centering
\includegraphics[width=\columnwidth]{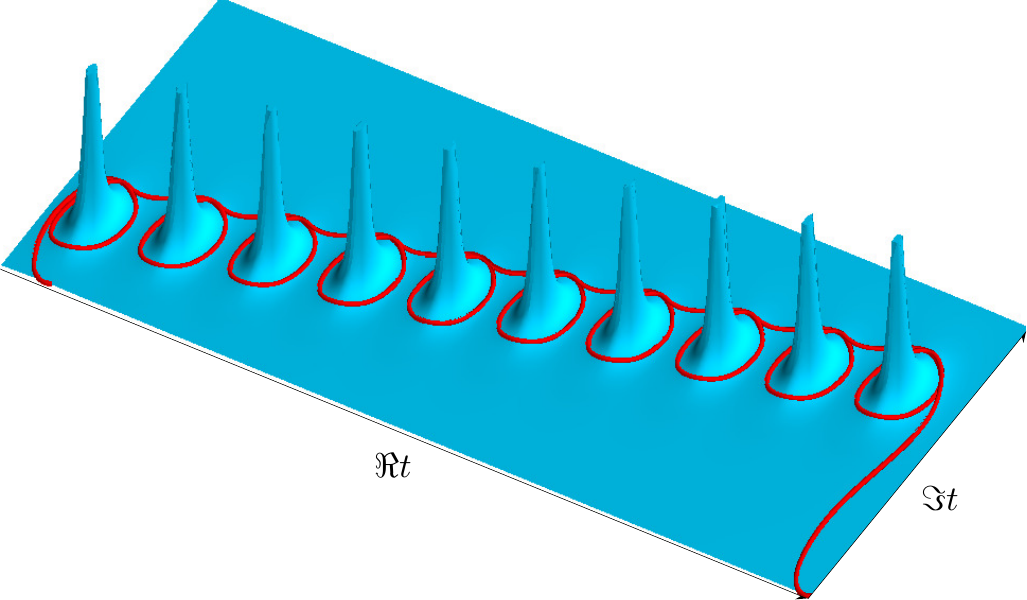}
\caption{\label{fig:GS_contour}Complex time integration contour for reconstructing the Coulomb ground state. Note how the contour circumnavigates the singularity times.}
\end{figure}

The total complex time contour thus consists of a series of loops around successive singularity times as shown in Fig.~\ref{fig:GS_contour}.
Note that the wavefunction is always reconstructed at purely real times.
The necessary excursions back to the real time axis are not shown in Fig.~\ref{fig:GS_contour}.

\begin{figure}[htp]
  \centering
\includegraphics[width=\columnwidth]{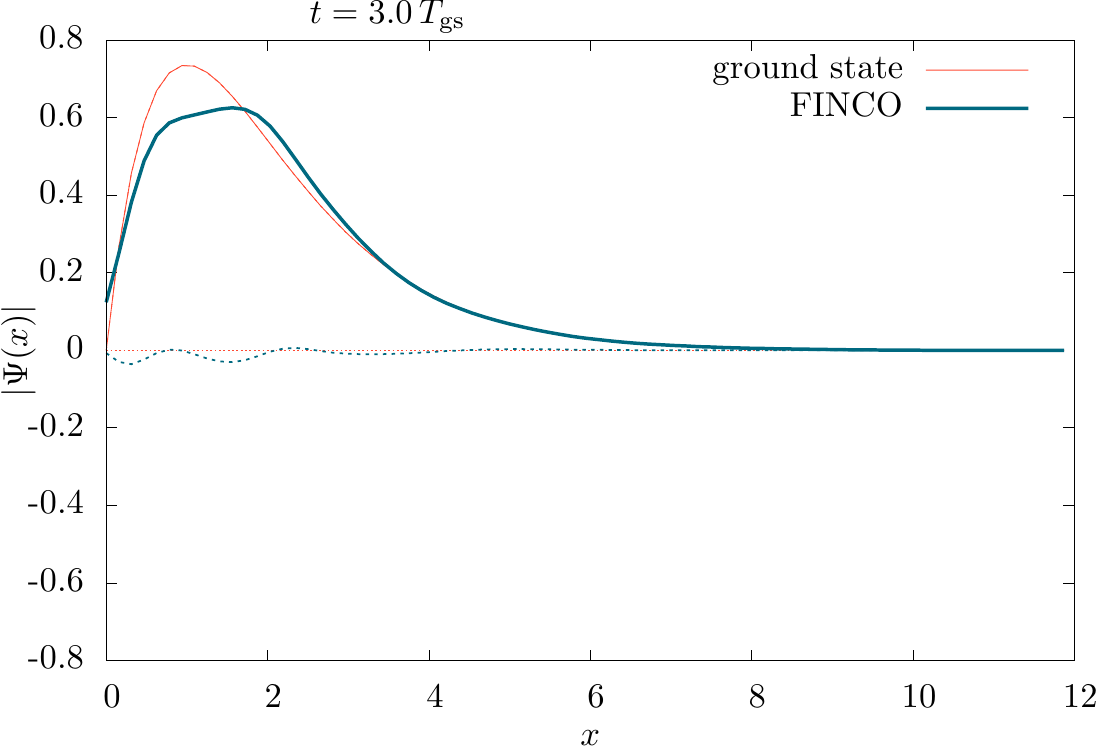}
\caption{\label{fig:GS_wavefunction}Coulomb ground state (thin red) and FINCO trajecory reconstruction at $t=3\cdot 4\pi$ with $\gamma=\frac{1}{2}$. ($\vert \Psi\vert$ thick blue, $\Im \Psi$, dashed blue)}
\end{figure}

The locations of singularity times depend on the trajectory initial conditions.
Propagating the entire manifold with suitably chosen contours yields the desired stationary ground state depicted in Fig.~\ref{fig:GS_wavefunction}, the only time-dependence being a coordinate-independent phase rotation.
Both the magnitude and phase of the wave function are in semiquantitative agreement with the exact quantum results.

\mysection{Step 2: Strong field dynamics of the Coulomb system}
Having described the construction of the initial wavefunction, we now introduce the time-dependent electric field.
The laser field is modeled in the dipole approximation and length gauge by the additional, time dependent potential $\tilde{V}(q,t)=e q F_0\sin(\omega t)$ with the field strength $F_0=7.35\cdot10^{-2}\unit{au}$, electronic charge $e$ and frequency $\omega=7.35\cdot10^{-2}\unit{au}$. The Keldysh parameter is $\gamma=1$.

\begin{figure}[htp]
  \centering
\includegraphics[width=\columnwidth]{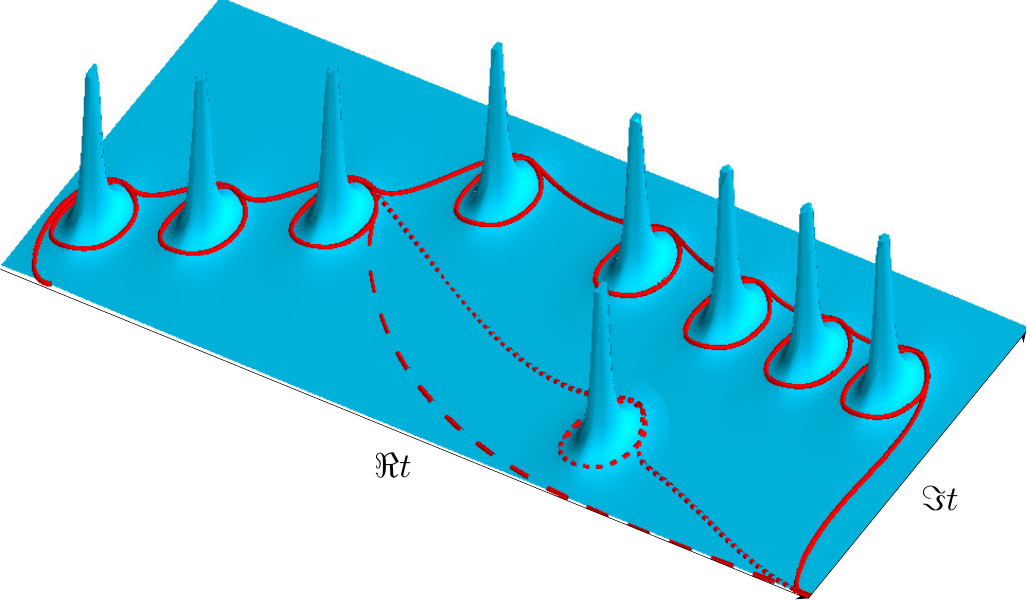}
\caption{\label{fig:SF_contours}Integration contours describing an initial stationary state as a strong external field is turned on. Different sections of the contour correspond to different processes: bound, core wave packet (solid), ionization (dashed) and recollision (dotted).
Note that singularity times have shifted with respect to the field free case Fig.~\ref{fig:GS_contour} and an additional singularity time is present.}
\end{figure}

The integration contour in the presence of the field is shown in Fig.~\ref{fig:SF_contours}. Note that the addition of $\tilde{V}(q,t)$ does not introduce additional singularities in complex space.
It does, however, shift the singularity times and introduces additional encounters with the Coulomb singularity.
The loops around the shifted singularity times correspond to orbiting trajectories that remain near the core of the nucleus, modified by the presence of the field \footnote{In practice, in order to avoid numerical inaccuracies in the ground state, we skip the initial field free loops and start with the field dressed loops at $t=0$.}.
However, the loops that circumnavigate the new singularity encounter correspond to a new kind of trajectory dynamics: ionization of the electron and subsequent recollision with the nucleus.
The ionizing and recolliding trajectories result from three distinct parts of a complex time contour :
\begin{inparaenum}
\item A series of singularity loops causing the trajectory to orbit the singularity at the origin.
Termination of the loop sequence initiates tunnel ionization.
\item
Straight line integration in complex time results in a field-mediated excursion from the nucleus.
Depending on the trajectory initial conditions, the trajectory will ionize or return to the core at a later time.
\item A loop around a new singularity time that is not present in the field free case induces recollision with the nucleus.
\end{inparaenum}

Each additional cycle of the field results in additional contributions to the emitted wavepacket.
These contributions are generated via contours analogous to those of the first ionization cycle, albeit with a varying number of initial loops.
E.g. the first ionization cycle begins after three initial loops whereas the second ionization cycle requires fifteen loops.
Note that the time of maximum field strength roughly coincided with three periods of phase rotation of the ground state and a full cycle of the field corresponds roughly to twelve periods of rotation of the ground state such that the second field maximum occurs after about fifteen periods of the ground state rotation.  
Terminating the initial loop series earlier or later yields trajectories that describe the same process but that disagree with the quantum results. The physical significance of these alternate contours is not yet understood.

\begin{figure}[htp]
  \centering
\includegraphics[width=\columnwidth]{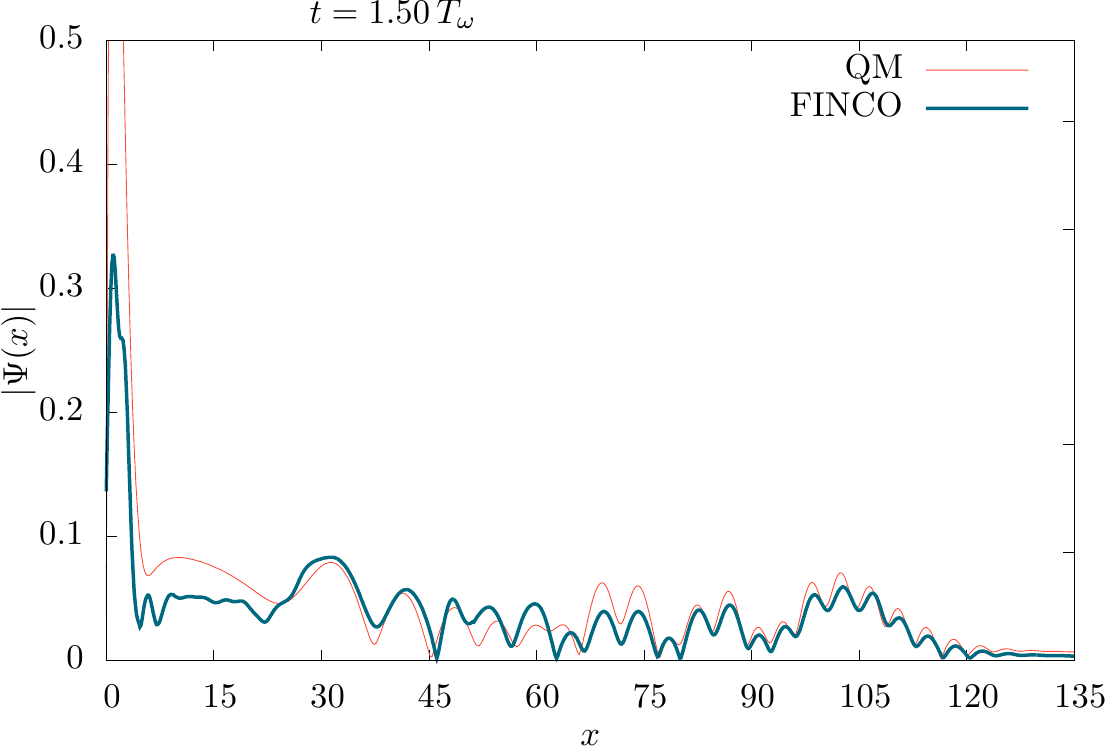}
\caption{\label{fig:SF_wavefunction}Coulomb ground state evolved under laser field influence: Split operator quantum solution (thin red) and FINCO trajecory reconstruction (thick blue)  at $t=1.5\cdot T_{\omega}\approx 10.2\cdot 4\pi$ with $\gamma=0.3$.
Included trajectories correspond to the core wave packet, first and third half cycle ionization and second half cycle recollision.}
\end{figure}

Adding the contributions from all distinct contours that make significant numerical contributions yields the wavefunction in Fig.~\ref{fig:SF_wavefunction}.
The ionized part of the wave function plotted for 1.5 periods of the field shows excellent agreement with the quantum result, including the complex nodal pattern resulting from the superposition of the three distinct contributions, one from each of the three half cycles of the field prior to the evaluation time.

As the laser field changes sign, the previously emitted part of the wavefunction is partially redirected towards the nucleus.
This recolliding wavefunction recombines with the ground state of the system, and can be computed from the trajectory based reconstruction.

\begin{figure}[htp]
  \centering
\includegraphics[width=\columnwidth]{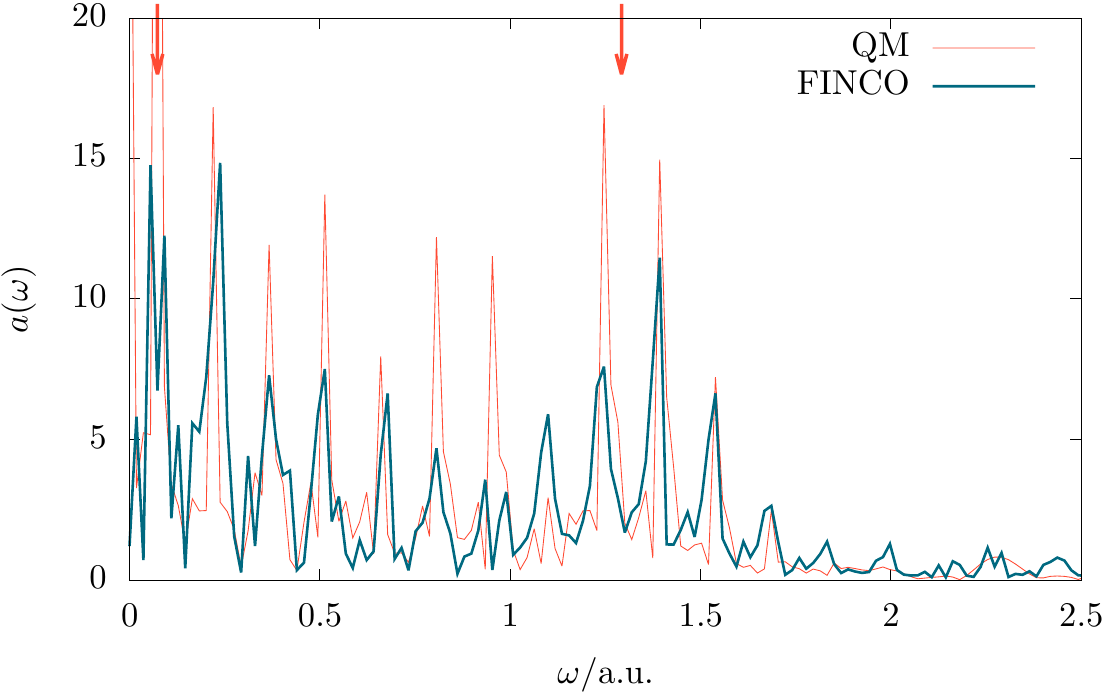}
\caption{\label{fig:HHG}High harmonic spectrum obtained from the Fourier transform of Eq.~\eqref{eq:dipoleacc} for four periods of the field. Quantum mechanical result (thin red) and FINCO (thick blue). Arrows indicate the fundamental frequency of the field and the corresponding expected cutoff at $3.17 U_{\rm p} +I_{\rm p}$. }
\end{figure}

\mysection{Step 3: Calculating the high harmonic spectrum}
The spectrum of the radiation emitted as the recolliding wavefunction recombines with the ground state of the system can be calculated
as the Fourier transform of the dipole acceleration computed in the form \cite{burnett_calculation_1992}
\begin{align}
\label{eq:dipoleacc}
  a(t)=\matrixel{\Psi(t)}{\ddot{x}}{\Psi(t)}=\int \di x \vert\Psi(x,t)\vert^2\frac{\partial V(x,t)}{\partial x}.\,
\end{align}
A comparison of the quantum and complex trajectory calculation is shown in Fig.~\ref{fig:HHG}.
The locations of the peaks as well as the plateau structure, including the cutoff, are reproduced very well.
Remaining deviations in the intensity of the peaks are due to errors in the reproduction of the core wavepacket.

Several comments about our numerical procedure. The FINCO wavefunction was reconstructed from four distinct classical processes: the first and third half cycle ionization and the second and fourth half cycle recollision.
The dominant regions in the initial manifold contributing significantly to each of these processes were selected manually.
Since the quantitative simulation of the core wavepacket is somewhat more involved, we added to the ionized and recolliding parts of the wavefunction, the Coulomb ground state with its time-dependent phase, $\Psi(x,t)\rightarrow \Psi(x,t)+\Psi_0(x)\e^{\frac{\imath}{2} t}$.

All of the data presented was computed from initial conditions in only the positive half of coordinate space, evolving on a single Riemann sheet of the potential $V(q)=\mp\frac{1}{q}$.
Instead of computing the negative half from the initially antisymmetric ground state $\Psi_0(x)$ evolving on the other sheet of the potential, the dipole acceleration was post-symmetrized according to $a(t)\rightarrow a(t)-a(t-\frac{\pi}{\omega})$.

\mysection{Summary}
To summarize, we have shown how strong field dynamics of the Coulomb system, including high harmonic generation, can be modeled using complex classical trajectories.
In this approach the spectrum emerges in the three following steps:
1) Trajectories orbit the singularity in the complex plane, generating the Coulomb ground state.
2) Under the influence of the strong field, a subset of the trajectory manifold ionizes, with a subset of the subset recolliding with the nucleus.
3) Recombination with the nuclear wavefunction yields high harmonic radiation.
There is no need for a separate tunnel ionization step as in the traditional three step model; it is included automatically in the second step.

The high harmonic spectrum reconstructed from the FINCO trajectories agrees well with the full quantum results.
With  the complex trajectories correctly describing tunnel ionization of the Coulomb ground state, a number of exciting future possibilities present themselves.

There has been significant debate in the literature whether the transition of the tunneling electrons through the barrier happens instantaneously or with a delay \cite{kaushal_nonadiabatic_2013,landsman_tunnelling_2014,zipp_probing_2014,pedatzur_attosecond_2015,torlina_interpreting_2015,ni_tunneling_2016}.
Since the complex trajectories used in the present work are rigorously derived from the Coulomb ground state without artificially imposing tunnel exit initial conditions, tunneling times may be read off directly. Note that each trajectory has its own independent tunneling time, allowing for transparent analysis and interpretation.

Aside from the manual selection of the significant regions of the initial manifold, the only approximation in the present approach is a truncation of an infinite hierarchy of EOMs at second order~\cite{goldfarb_bohmian_2006}, corresponding to a saddle point evaluation of the path integral \cite{schiff_path-integral_2011}.
Since no assumption is made on physical parameters such as the strength of the field, we expect that the complex trajectory method presented here will provide mechanistic insight into a broad range of strong field effects in atomic and molecular systems.

\vspace{1EM}
{\it Acknowledgements}  We thank the organizers and participants of the Advanced Study Group ``Semiclassical Methods: Insight and Practice in Many Dimensions'' and the Focus Workshop ``Quantum and Semiclassical Trajectories'', both held at the MPIPKS of Dresden, for many stimulating discussions. Financial support from the Israel Science Foundation (1094/16) and the German-Israeli Foundation for Scientific Research and Development (GIF) are gratefully acknowledged.

\appendix
\renewcommand{\theequation}{\thesection.\arabic{equation}}

  \section{Equations of motion for stability matrix elements}
  \label{sec:stability}

The stability matrix elements $M_{ab}=\frac{\partial a_t(q_0)}{\partial b_0(q_0)}$ with $a,b\in\{p,q\}$  obey the following EOMs
            \begin{align}
            \dot{M}_{\rm pp} &= -V''(q)M_{\rm qp} & \dot{M}_{\rm pq} &= -V''(q)M_{\rm qq} \\
    \dot{M}_{\rm qp} &= \frac{1}{m}M_{\rm pp} & \dot{M}_{\rm qq} &= \frac{1}{m}M_{\rm pq}\,.
  \end{align}

\restoretitle{A high harmonics three step model in complex space and complex time}

\bibliographystyle{apsrev}
\dontignore{\end{document}}

\section{Elimination of Stokes divergences}
\label{sec:stokes}
In principle, the complex trajectories resulting from the relation Eq.~\eqref{eq:manifold} represent saddles of the path integral of the system \cite{schiff_path-integral_2011} obtained through deformation of the integration contour into the complex plane.
However, while this saddle point approximation yields the correct classical limit, straightforward integration of the entire manifold of initial conditions includes saddles that correspond to deformations of the integration contour beyond singular points in the complex plane.
These initial condition are physically invalid, resulting in a so called Stokes divergence \cite{stokes_discontinuity_1857,stokes_discontinuity_1902} of the coefficients $\phi(q_0)\e^{\sigma(q_0)}$.

A procedure for determining whether a particular trajectory is in a Stokes divergent region was introduced in Ref.~\cite{koch_communication:_2018}.
In brief, we locate caustics in the manifold as roots of Eq.~\eqref{eq:DxiDnu}.
For each caustic $q_0^{\ast}$ we define the phase space caustic Stokes expansion (PCSE) in terms of the initial coordinate $q_0$
\begin{align}
\label{eq:taylor_F_xi}
F(q_0)=\left[\frac{1}{3}\sigma^{(3)}(q_0^{\ast})+\sigma^{(2)}(q_0^{\ast})\varrho\right]\tilde{q}_0(q_0)^{3}+\order{\tilde{q}_0(q_0)^5}\,,
\end{align}
where $\tilde{q}_0(q_0){=}{\pm}\left(\frac{\xi(q_0)}{\xi^{(2)}(q_0^{\ast})}\right)^{\frac{1}{2}}$ with the sign of $\tilde{q}_0$ chosen such that $|\tilde{q}_0(q_0)-q_0|$ is minimal and $\varrho=\frac{1}{3}\sigma^{(3)}(q_0^{\ast})+\sigma^{(2)}(q_0^{\ast})\frac{-\xi^{(3)}(q_0^{\ast})}{3\xi^{(2)}(q_0^{\ast})}$. Parenthesized superscripts denote differentiation with respect to $q_0$ at $q_0^{\ast}$.
The higher derivatives $\xi^{(i)}(q_0^{\ast}),\sigma^{(i)}(q_0^{\ast}),i\in\{2,3\}$ are computed via finite differencing with trajectories near $q_0^{\ast}$.

The $q_0$ manifold is split into six sectors along the anti Stokes lines $\Re F(q_0)=0$; the sector that contains the Stokes line $\Im F(q_0)=0$ along which $\Re\sigma(q_0)>0$ diverges is removed.
All trajectory contributions $\phi(q_0)\e^{\sigma(q_0)}$ in the two adjacent sectors are multiplied by the Berry factor $S(F(q_0))=\erf \left\{\frac{\Im F(q_0)}{\sqrt{2\Re F(q_0)}}\right\}$ where $\erf(\tau){=}\frac{1}{\sqrt{\pi}}\int_{-\infty}^\tau\di{s}\e^{-s^2}$ is the error function \cite{berry_uniform_1989}.
For details on this procedure and derivations of the expressions given above see Ref.~\cite{koch_communication:_2018}.